\documentstyle[12pt, a4]{article}
\begin{document}
\author{Be\c{s}ire G\"{o}n\"{u}l, B\"{u}lent G\"{o}n\"{u}l, Dilek Tutcu 
and Okan \"{O}zer
\and Department of Engineering Physics,\and
University of Gaziantep, 27310 Gaziantep-T\"{u}rkiye}
\title{Supersymmetric approach to exactly solvable systems
with position-dependent effective masses}
\maketitle
\begin{abstract}
We discuss the relationship between exact solvability of the
Schr\"{o}dinger equation with a position-dependent mass and the
ordering ambiguity in the Hamiltonian operator within the frame of
supersymmetric quantum mechanics. The one-dimensional
Schr\"{o}dinger equation, derived from the general form of the
effective mass Hamiltonian, is solved exactly for a system with
exponentially changing mass in the presence of a potential with
similar behaviour, and the corresponding supersymmetric partner
Hamiltonians are related to the effective-mass Hamiltonians
proposed in the literature.
\end{abstract}


\section{Introduction}
The study of quantum systems with position-dependent effective
masses has been the subject of much activity in recent years. The
Schr\"{o}dinger equation with nonconstant mass  provides an
interesting and useful model for the description of many physical
problems. The most extensive use of such an equation is in the
physics of semiconductor nanostructures. This field has arisen due
to the impressive devolepment of sophisticated technologies of
semiconductor growth, like molecular beam epitaxy, which made it
possible to grow ultrathin semiconductor structures, with very
prominent quantum effects \cite{bastard1}. The motion of electrons
in them may often be described by the envelope function
effective-mass Schr\"{o}dinger equation, where the material
composition- (i.e., the position-) dependent effective mass of
electrons replaces the constant particle mass in the conventional
Schr\"{o}dinger equation. The most popular of these structures is
the semiconductor quantum well, and the Schr\"{o}dinger equation
here is effectively one-dimensional. Another instance where such
an equation is employed, this time three-dimensional and with
spherical symmetry, is in the pseudopotential-theory-based density
functional calculations in solids: to reduce the computational
load, model pseudopotentials with position-dependent electron mass
which replace nonlocal pseudopotentials have been considered
\cite{foulkes}.

Since the momentum and the mass operators no longer commute in
case of spatially varying mass, a question concerning the correct
form of the kinetic energy operator of the generalized Hamiltonian
has arisen. This problem of ordering ambiguity is a long standing
one in quantum mechanics, see for instance the excellent critical
review by Shewell \cite{shewell}. There are many examples of
physically important systems, for which such ambiguity is quite
relevant. For instance we can cite the problem of impurities in
crystals \cite{luttinger,wannier,slater}, the dependence of
nuclear forces on the relative velocity of the two nucleons
\cite{rojo,razavy}, the minimal coupling problem in systems of
charged particles interacting with magnetic fields \cite{landau},
and more recently the study of semiconductor heterostructures
\cite{bastard1,weisbuch,harrison}.

Notwithstanding, taking into account the spatial variation of the
semiconductor type, some effective Hamiltonians are proposed with
a spatially dependent mass for the carrier [12-17], and many
authors have been trying to determine the correct Hamiltonian
phenomenologically. In this article we try to circumvent the
problem of ambiguity by presenting a scheme to obtain
unambiguously the Schr\"{o}dinger equation with spatially varying
particle mass, which makes clear the link between possible choices
of the kinetic energy operator for quantum systems with
position-dependent effective mass, within the frame of
supersymmetric quantum mechanics \cite{cooper1}. The strategy here
is to tackle the problem with a fundamental point of view, ie.,
without using a particular form of the effective potential. Within
this framework and using the supersymmetric formalism we will show
that one can arrive at a conceptually consistent result for
exactly solvable systems.

The application of supersymmetry ideas to nonrelativistic quantum
mechanics has provided a deeper understanding of analytically
solvable Hamiltonians, as well as a set of powerful approximate
schemes for dealing with problems admitting no exact solutions.
The concept of shape invariance \cite{gendenshtein} has played a
fundamental role in these developments. The aims of the present
work are to consider the application of the supersymmetric
approach to quantum systems with position-dependent mass and to
extend the concept of shape invariant potentials to the
nonconstant mass scenario to see clearly the relation between the
effective-mass potentials existing in the literature and
supersymmetric partner potentials.

The paper is organized as follows. In section 2 we provide a brief
review of the Schr\"{o}dinger equation for systems with
position-dependent effective mass. Section 3 deals with the
exact solvabilty of a system involving a particle with a spatially
dependent mass in an arbitrary potential well. The application of
supersymmetric approach to this system, together with the use of
shape invariance concept to deduce the correct operator ordering
for the Hamiltonian, are also studied in this section. Finally,
some conclusions are drawn in section 4.

\section{Generalized Schr\"{o}dinger equation}
We start this section by defining a quite general Hermitian
effective Hamiltonian for the case of a spatially varying mass
which will be denoted by $m~[=m(x)]$ throughout the present work. 
In general, the Hamiltonian
proposed by von Roos \cite{von} is used,
\begin{equation}
H_{VR}=\frac{1}{4}\left[m^{\alpha}~pm^{\beta}~p~m^{\gamma}
+m^{\gamma}~p~m^{\beta}~p~m^{\alpha}\right]~,
\end{equation}
but to accomodate the possibility of including the case of the
Weyl ordering \cite{saborges} in a more evident way, we will use
an effective Hamiltonian with four terms given by \cite{desouza}
\begin{equation}
H=\frac{1}{4(a+1)}\left\{a~[m^{-1}p^{2}+p^{2}m^{-1}]+
m^{\alpha}~pm^{\beta}~pm^{\gamma}
+m^{\gamma}pm^{\beta}pm^{\alpha}\right\}~.
\end{equation}
In both cases, Eqs. (1) and (2), there is a constraint over the
parameters such that $\alpha+\beta+\gamma=-1$.

The one-dimensional time-independent Schr\"{o}dinger equation 
reads
\begin{equation}
-\left(\frac{\hbar^{2}}{2m}\right)\frac{d^{2}\psi}{dx^2}+
\frac{\hbar^{2}}{2}\left(\frac{m'}{m^2}\right)\frac{d\psi}{dx}+
[U_{\alpha\gamma a}+V]\psi=E\psi~,
\end{equation}
where $U_{\alpha\gamma a}$
involving all the ambiguity is
\begin{equation}
U_{\alpha\gamma
a}(x)=-\frac{\hbar^{2}}{4m^{3}(a+1)}\left[(\alpha+\gamma-a)m~m''+
2(a-\alpha\gamma-\alpha-\gamma)m'^{2}\right]~,
\end{equation}
in which the first and second derivatives of $m(x)$with respect to
$x$ are denoted by $m'$and $m''$. It is clear that the effective
potential is the sum of the real potential profile
$V(x)$ and the modification  $U_{\alpha\gamma a}$ emerged from the
location dependence of the effective mass. A different Hamiltonian
leads to a different modification term, see Table 1.

It is curious to note that all the ambiguity is in the
$U_{\alpha\gamma a}$ term, and that it can be eliminated by
imposing some convenient constraints over the ambiguity
parameters, namely
\begin{equation}
\alpha+\gamma-a=0,~~~~a-\alpha\gamma-\alpha-\gamma=0,
\end{equation}
which have two equivalent solutions,(i) $\alpha=0$ and $a=\gamma$,
or (ii) $a=\alpha$ and $\gamma=0$. In this case the effective
Schr\"{o}dinger equation will not depend on the ambiguity
parameters, but will contain a first order derivative term. In the
next section, we will be interested in getting exact solutions of
the resulting equation for a particular potential, and trying to
get some information about the proposed orderings appearing in the
literature.

\section{An exactly solvable system}
The interest in exactly solvable problems in quantum physics has
increased sharply in the last few years. This is concerned, of
course, with the fact that the description of the behaviour of
nonconservative physical systems is usually very complicated, but
in some cases such systems can be modelled by means of quite a
simple Hamiltonian, which leads to standard problems of quantum
mechanics.

Starting with the Schr\"{o}dinger equation in Eq. (3) and making a
new definition for the wave function
\begin{equation}
\psi(x)=m^{1/2}\varphi(x),
\end{equation}
one gets a differential equation in a more familiar form
\begin{equation}
-\frac{\hbar^{2}}{2m}\frac{d^{2}\varphi}{dx^{2}}+\left(U_{eff}-E\right)\varphi=0,
\end{equation}
with a new effective potential defined through
\begin{equation}
U_{eff}(x)=V(x)+U_{\alpha\gamma a}(x)+
\frac{\hbar^{2}}{4m}\left[\frac{3}{2}\left(\frac{m'}{m}\right)^{2}-
\frac{m''}{m}\right]~.
\end{equation}
To demonstrate the simplicity of the present method, we consider
here a particular case, which has been recently studied
\cite{desouza}, where one have exact solution for the above
equation. That is a particle with exponentially decaying or
increasing mass in the presence of a potential with similar
behaviour,
\begin{equation}
m(x)=m_{0}~e^{cx},~~~~V(x)=V_{0}~e^{cx},
\end{equation}
where $m_{0}$ is a constant mass. This problem is often
encountered in the calculation of confined energy states for
carriers in semiconductor quantum well structures under the
envelope-function and the effective-mass approximations where the
effective mass of a carrier is spatially dependent on the graded
composition of the semiconductor alloys used in the barrier and
the well region of the microstructures.

Multiplying each term in Eq. (7) by $m(x)$ and dividing by
$m_{0}$, one arrives at an usual Schr\"{o}dinger equation for the
system of interest,
\begin{equation}
-\frac{\hbar^{2}}{2m_{0}}\frac{d^{2}\varphi}{dx^{2}}+
\left(V_{0}~e^{2cx}-E~e^{cx}\right)\varphi=\varepsilon\varphi~,
\end{equation}
where
\begin{equation}
\varepsilon=\frac{\hbar^{2}}{m_{0}}(q-c^{2}/8),~~
q=\frac{c^{2}}{4(a+1)}(a-2\alpha\gamma-\alpha-\gamma).
\end{equation}
Note that Eq. (10) corresponds to a Schr\"{o}dinger equation for a
particle with constant mass under the influence of the Morse
potential \cite{morse}. It is thus clear that if one knows the
spectral properties of the constant-mass Schr\"{o}dinger equation
of any potential, one then can readily obtain a corresponding
potential for the effective-mass Schr\"{o}dinger equation with
identical spectral properties. Proceeding with the well known
energy spectrum of the Morse potential,
\begin{equation}
-\frac{\hbar^{2}c^{2}}{2m_{0}}\left[\frac{\sqrt{2m_{0}V_{0}}}{\hbar
c}-\left(n+\frac{1}{2}\right)\right]^{2}=
\frac{\hbar^{2}}{m_{0}}\left(q-\frac{c^{2}}{8}\right),
\end{equation}
from which we find the energy
spectrum of the effective potential appearing in Eq. (7), that 
is 
\begin{equation}
E_{n}=\hbar~c~
\sqrt{\frac{V_{0}}{2m_{0}}}~[2n+1+\nu(\alpha,\gamma,a)].
\end{equation}
where the ordering term $\nu(\alpha,\gamma,a)$ is
\begin{equation}
\nu(\alpha,\gamma,a)=\sqrt{1-\frac{2}{a+1}(a-2\alpha\gamma-\alpha-\gamma)}~.
\end{equation}

One can now study the effect of using some of the orderings
appearing in the literature. Considering Table 1, it is not
difficult to see that the ambiguous term $\nu$ is zero for the
effective Hamiltonians in Refs. \cite{zhu,li,saborges}, although
they have different orderings, while $\nu=1$ for the
BenDaniel-Duke Hamiltonian \cite{bendaniel}. For both cases
($\nu=0,1$), the corresponding Hamiltonians have exactly the same
spectra except for the fact that the Hamiltonians in Refs.
\cite{zhu,li,saborges} have one bound state more than the
BenDaniel-Duke Hamiltonian.  Thus, they can be treated as the
supersymmetric partner Hamiltonians \cite{cooper1} which is the
subject of the next section. Furthermore, one ends with a complex
energy for the orderings proposed in the literature
\cite{gora,bastard2}, which could be possibly discarded due to the
physically unacceptable energies. This makes clear that
unacceptable physics consequences may occur unless specific
choices are made in the Hamilton operator ordering for a system
undertaken.

As we are dealing with a confined particle system, one may 
wish also to confirm Eq. (13) by mapping the Morse potential 
onto harmonic oscillator system, which seems more realible than 
the Morse oscillator for the system of interest. For this purpose, 
we invoke the change in the variable as well as in the wave function,
\begin{equation}
x=\ln y^{2/c}~~,\varphi(x)=\sqrt{2/c}~y^{-1/2}~F(y)~.
\end{equation}

This reduces Eq. (10) to an equivalent Schr\"{o}dinger equation
\begin{equation}
-\frac{\hbar^{2}}{2m_{0}}\frac{d^{2}F}{dy^{2}}+
\left[\frac{4V_{0}}{c^{2}}y^{2}-
\left(\frac{\hbar^{2}c^{2}+32m_{0}\varepsilon}{8m_{0}c^{2}y^{2}}\right)
\right]F=\frac{4E}{c^{2}}F~,
\end{equation}
in a more familiar form involving a harmonic oscillator potential
with centripetal barrier. In contrast to Eq. (10), which contains
a variable parameter $\varepsilon$ representing the Morse
oscillator energy, we have a constant term on the right-hand side
of Eq. (16), and the energy term ($\varepsilon$) is contained in
the effective potential parameter. Thus Eq. (16) may be looked
upon as the radial Schr\"{o}dinger
equation with a fixed energy but with a variational angular
momentum quantum number. From which one can easily arrive at Eq. (13), 
which clarifies that both treatments (Morse and harmonic oscillator 
mapping) are equivalent to each other.

In the following section, we will focus our attention 
on how to apply the supersymmetric
quantum mechanical formalism to the system under consideration in
order to clear out the hidden relation between the effective
Hamiltonians proposed in the literature for the spatially dependent
mass.

\subsection{supersymmetric approach}
The problem of generating isospectral potentials in quantum
mechanics has been considered for more than 50 years, but recently
the research efforts on this topic have been considerably
intensified. A new field, supersymmetric quantum mechanics
(SUSYQM), devoted to this class of problems has emerged, which
deals with pairs of Hamiltonians that have the same energy
spectra, but different eigenstates. A number of such pairs of
Hamiltonians share an integrability condition called shape
invariance \cite{gendenshtein}. Although not all exactly solvable
problems are shape invariant \cite{cooper2}, shape invariance,
especially in its algebraic formulation, is a powerful technique
to study exactly integrable systems which have always been at the
centre of attention in physics and mathematics.

It would be interesting therefore to extend the SUSYQM to handle
cases with position-dependent mass. Recently supersymmetric
techniques have been applied to obtain exact solutions of
Schr\"{o}dinger equation with nonconstant mass
\cite{milanovic,plastino}. Using the spirit of these works, here
we generalize the supersymmetric formalism for the problem
considered in the previous section. All considerations are made
for the one-dimensional Schr\"{o}dinger equation.

Proceeding as in the case of constant mass, we introduce a
superpotential $W(x)$ and the associated pair of operators $A$ and
$A^{+}$ defined by
\begin{equation}
A\psi=\frac{\hbar}{\sqrt{2m}}\frac{d\psi}{dx}+W\psi~~~~,~~~~
A^{+}=-\frac{d}{dx}\left(\frac{\hbar\psi}{\sqrt{2m}}\right)+W\psi~.
\end{equation}
Notice that, due to the position dependence of the mass, $d/dx$
and $\hbar/\sqrt{2m}$ do not commute anymore. Within the framework
of SUSYQM, the first partner Hamiltonian reads
\begin{equation}
H_{1}=A^{+}A=-\frac{\hbar2}{2m}\frac{d^{2}}{dx^{2}}-
\left(\frac{\hbar^{2}}{2m}\right)'\frac{d}{dx}+ \left[
W^{2}-\left(\frac{\hbar W}{\sqrt{2m}}\right)'\right]~,
\end{equation}
where the prime denotes the first derivative with respect to the
variable $x$.

At this stage, we propose an ansatz for the superpotential, 
\begin{equation}
W(x)=\frac{\hbar~c}{8~m_{0}}~\sqrt{2m}-\frac{\hbar~c}{2\sqrt{2m}}~,
\end{equation}
and note that the Hamiltonian in Eq. (18) found
via the supersymmetric formalism corresponds to the specific effective-mass
Hamiltonians in Eq.(3) with real potential profile $V(x)$. Explicitly, 
these are the Zhu and Kroemer Hamiltonian 
($a=0, \alpha=\gamma=-1/2$) \cite{zhu},
the Li and Kuhn Hamiltonian ($a=\alpha=0, \beta=\gamma=-1/2$) \cite{li},
and the Weyl Hamiltonian ($a=1, \alpha=\gamma=0$) \cite{saborges},
in which the effective potentials (for which $\nu(\alpha,\gamma,a)=0$) 
may be expressed in the supersymmetric form,
\begin{equation}
V_{eff}(x)=V_{0}~e^{cx}+U_{\alpha\gamma a}(x)=
W^{2}-\left(\frac{\hbar W}{2m}\right)'~~.
\end{equation}
This justifies the operator ordering in the Hamiltonian used by
many authors in the rough calculation of the confinement states in
quantum well structures in the effective mass scheme, as the first
partner leads to the exactly solvable Hamiltonian systems within
the frame of SUSYQM.

The associated supersymmetric partner Hamiltonian, for the case
$\nu(\alpha,\gamma,a)=1$ which describes the BenDaniel-Duke
effective mass Hamiltonian ($a=\alpha=\gamma=0$) \cite{bendaniel},
is
\begin{equation}
H_{2}=AA^{+}=-\frac{\hbar^{2}}{2m}\frac{d^{2}}{dx^{2}}-
\left(\frac{\hbar^{2}}{2m}\right)'\frac{d}{dx}+
\left[W^{2}-\left(\frac{\hbar W}{\sqrt{2m}}\right)'+ \frac{2\hbar
W}{\sqrt{2m}}-\frac{\hbar}{\sqrt{2m}}
\left(\frac{\hbar}{\sqrt{2m}}\right)''\right]
\end{equation}
We see that the two supersymmetric partner Hamiltonians $H_{1}$
and $H_{2}$ describe particles with the same effective
mass-spatial dependence, but in different potentials. The second
partner potential corresponding the BenDaniel-Duke effective
potential does not incorporate the ambiguity term,
\begin{equation}
V_{BDD}(x)=V_{0}~e^{cx}=W^{2}-\left(\frac{\hbar
 W}{\sqrt{2m}}\right)'+ \frac{2\hbar
W}{\sqrt{2m}}-\frac{\hbar}{\sqrt{2m}}
\left(\frac{\hbar}{\sqrt{2m}}\right)''~,
\end{equation}
where the double prime denotes the second derivative with respect
to $x$.

There is a correspondence between the energy eigenvalues of the
isospectral Hamiltonians $H_{1}$ and $H_{2}$, although they have
different effective potentials. The energy of the $n$th bound
state of $H_{1}$ coincides with that of the $(n-1)$th bound state
of $H_{2}$, which is the case expressed through Eq. (13). The
ground state of $H_{1}$ has no associated state of $H_{2}$. The
reader is referred to Ref. \cite{cooper1} for a comprehensive
review of  the supersymmetric quantum mechanics.

For an exactly solvable system it is thus obvious that a proper
choice of the effective mass and the corresponding superpotential 
would impose the elimination of the
supersymmetric partner potential that involves all the ambiguity. 
This makes clear that which choice of the effective mass Hamltonians 
proposed in the literature is physically acceptable for the system 
undertaken.

Finally we note that, from the relation between the superpotential and the
ground state wave function of $H_{1}$,
\begin{equation}
W(x)=-\frac{\hbar}{\sqrt{2m}}\frac{d\psi_{1}^{n=0}/dx}{\psi_{1}^{n=0}},
\end{equation}
it is straightforward to solve $\psi_{1}^{n=0}$, 
which satisfies the required boundary
conditions dictated from the conservation of current through the
envelope functions of heterostructures.

In sum, considering the definition of the effective potential
which is the summation of a real potential $V(x)$ and a term
$U_{\alpha\gamma a}(x)$ resulting from the mass dependence on the
location, and bearing in mind that the effective potential relies
on the Hamiltonian utilized, we have connected the deviation from
the real potential, due to the ordering ambiguity, to the
supersymmetric partner Hamiltonians. It is found that the operator
ordering in the kinetic energy operator to be the same as that
endorsed by recent authors [27, and the references therein] for
the calculation of confined states in semiconductor
microstructures under the simplified effective-mass and
envelope-function, with a correctional term $U_{\alpha\gamma
a}(x)$ that is proportional to the derivatives of the mass profile
$m(x)$.

\section{Concluding remarks}
In this work we have first discussed the problem of solvability
and ordering ambiguity in quantum mechanics, as the form of the
effective mass Hamiltonian has been a controversial subject due to
the location dependence of the effective mass.

It was shown through a particular example that exact solutions
could be used as a kind of guide, at least, restrict the possible
choices of ordering. The principal idea is to suppose that once
one have found the ordering without ambiguity for a given
potential or class of potentials, that ordering should be extended
to remaining physical potentials.

Our work has also made clear that the Hamiltonian proposed by Li
and Kuhn \cite{li} is in fact equivalent to that coming from the
Weyl ordering \cite{saborges}, as can be easily checked from 
Eqs.(4, 21, 24) the effective potentials and consequently from Eqs.
(13,14) the energy spectra of both these orderings are equal. In
addition, though the ordering in the Zho and Kroemer effective
Hamiltonian operator \cite{zhu} is different than those of Li and
Khun, and that of Weyl, we have arrived at the same result in each
case within the frame of both, the standard quantum theory and
supersymmetric quantum mechanics. Hence, we remark that when the
ambiguity term is linear in the momentum, contrary to what is
usually believed, there is no ambiguity in fact. Thus, any
Hermitian construction of the quantum Hamiltonian will be
necessarily equivalent to that due to any other, and consequently
nonambiguous. This observation supports the recent work of Dutra
and Almeida \cite{desouza}.

On the basis of the supersymmetric ideas we have generalized the
concept of shape invariance to the nonconstant scenario and shown
that an appropriate choice of the potential and  mass variation
with the position makes clear the link between the effective
potentials and the ordering ambiguity. The bound state spectra of
systems with effective mass are relavant in many areas  such as
the study of nuclei \cite{ring} and metal clusters \cite{puente}.
However, in other fields such as for instance electronic
properties of semiconductors, one is interested in the properties
of quantum systems with nonconstant mass endowed with continuum
spectra [1,10,11]. The supersymmetric formalism developed here can
also be useful tool for the treatment of such kinds of problems
and it may be possible to generalize our results to such problems.
Such an investigation will be deferred to a later publication

As a final remark, an exponentially changing mass and potential
have been considered in this paper, and it has been shown that the
energy levels could be redefined in such a way that the ordering
ambiguity disappears. However, we should note that the exact
solvability does depend upon both the form of the potential and
the change in mass with the position. For instance, as was
discussed in \cite{von}, this cannot be done for the consideration
of a quadratically growing mass in a singular potential field.
However, we believe that in addition to its practical
applications, the study of quantum mechanical systems with a
position-dependent mass within the frame of the present technique
will raise many interesting conceptual problems of a fundemantal
nature. In particular the method should find wide applications in
the study of quasi-exactly solvable \cite{boschi} and
conditionally-exactly solvable \cite{desouza2} systems with
nonconstant masses. Along this line the works are in progress.

\hspace{1.0in}

\newpage\

\newpage
\begin{table}[t]
\caption{\label{TABLE I}
        { The special cases of the effective Hamiltonian in Eq. 2.
        }}
\vspace{5mm}
\begin{center}
\begin{tabular}{ccccc}
\hline \hline
\\
Hamiltonian&a&$\alpha$&$\beta$&$\gamma$
\\
\hline
\\
Ref. [12]&0&0&-1&0\\ \\ Ref. [13,14]&0&-1&0&0\\ \\ Ref.
[15]&0&-1/2&0&-1/2\\ \\ Ref. [16]&0&0&-1/2&-1/2\\ \\ Ref.
[21]&1&0&0&0\\ \hline
\end{tabular}
\end{center}
\end{table}

\end{document}